**Percolating Superconductivity in Air-Stable Organic-Ion Intercalated MoS$_2$**

*Jose M. Pereira, Daniel Tezze, Iris Niehues, Yaiza Asensio, Haozhe Yang, Lars Mester, Shu Chen, Felix Casanova, Alexander M. Bittner, Maider Ormaza, Frederik Schiller, Beatriz Martín-García, Rainer Hillenbrand, Luis E. Hueso[*], Marco Gobbi[*]*

J. M. Pereira, D. Tezze, Dr. I. Niehues, Y. Asensio, Dr. H. Yang, Dr. L. Mester, Dr. S. Chen, Prof. F. Casanova, Prof. A. M. Bittner, Dr. B. Martín-García, Prof. R. Hillenbrand, Prof. L. E. Hueso, Dr. M. Gobbi
CIC nanoGUNE BRTA, San Sebastian, Basque Country 20018, Spain.
E-mail: l.hueso@nanogune.eu; m.gobbi@nanogune.eu

Dr. L. Mester
Sales Application Engineer, Attocubes systems AG, Munich, Bavaria 85540, Germany

Prof. F. Casanova, Prof. A. M. Bittner, Prof. R. Hillenbrand, Prof. L. E. Hueso, Dr. M. Gobbi
IKERBASQUE Basque Foundation for Science, Bilbao, Basque Country 48009, Spain.

Dr. M. Ormaza
Departamento de Polímeros y Materiales Avanzados: Física, Química y Tecnología, Facultad de Químicas UPV/EHU, Apartado 1072, 20080 Donostia-San Sebastián, Spain

Dr. F. Schiller, Dr. M. Gobbi
Centro de Física de Materiales (CFM-MPC) Centro Mixto CSIC-UPV/EHU
San Sebastián, Basque Country 20018, Spain.

Keywords: MoS$_2$, Superconductivity, Charge Density Wave, Organic-ion Intercalation, Electrochemical Intercalation, tetraethylammonium bromide (TEAB), cetyltrimethylammonium bromide (CTAB)

When doped into a certain range of charge carrier concentrations, MoS$_2$ departs from its pristine semiconducting character to become a strongly correlated material characterized by exotic phenomena such as charge density waves or superconductivity. However, the required doping levels are typically achieved using ionic-liquid gating or air-sensitive alkali-ion intercalation, which are not compatible with standard device fabrication processes. Here, we report on the emergence of superconductivity and a charge density wave phase in air-stable organic cation intercalated MoS$_2$ crystals. By selecting two different molecular guests, we show that these correlated electronic phases depend dramatically on the intercalated cation, demonstrating the potential of organic ion intercalation to finely tune the properties of 2D materials. Moreover, we find that a fully developed zero-resistance state is not reached in few-nm-thick flakes, indicating the presence of three-dimensional superconductive paths which are severed by the mechanical exfoliation. We ascribe this behavior to an inhomogeneous charge carrier distribution, which we probe at the nanoscale using scanning near-field optical microscopy. Our results establish organic-ion intercalated MoS$_2$ as a platform to study the emergence and modulation of correlated electronic phases.

## 1. Introduction

Van der Waals superconductors that can be isolated at the atomically thin limit represent a rich playground to study phase transitions in reduced dimensions[1–4] and explore the interplay between different correlated electronic phases,[1,2,5,6] such as superconductivity and charge density waves (CDWs). Moreover, these layered superconductors are emerging as a crucial material platform for quantum computing[7,8] as they offer the possibility to dramatically scale down the footprint of superconductive qubits when interfaced with ultrathin 2D dielectrics.[9,10] While superconductivity naturally occurs in some layered compounds such as NbSe$_2$[2,3,11] and 2H-TaS$_2$,[4] charge carrier doping leading to electron densities in the order of $10^{14}$ carriers/cm$^2$ per layer can drive other van der Waals materials into a superconducting state. For instance, exfoliated flakes of MoS$_2$, which is a semiconductor in its pristine state,[12,13] display a superconductive transition when exposed to the extreme electric fields generated by ionic liquid gating.[14–16] Another approach to achieving similar doping effects in layered compounds is the insertion of ions in their interlayer van der Waals (vdW) gap, a technique commonly known as intercalation.[17–19] This approach leads to dramatic changes in the electrical,[20,21] optical,[22] and magnetic properties[23,24] of the host layered compound, as extensively reported for bulk crystals[25,26] and more recently for exfoliated flakes.[22,27]

In the case of bulk MoS$_2$, the intercalation of alkali metals results in the emergence of phenomena such as superconductivity[25,26,28] and CDWs.[29] However, alkaline and alkaline-earth intercalated compounds offer poor ambient stability, complicating their exfoliation and thus their integration in nanoscale devices.[30] Recently, it has been shown that MoS$_2$ and other transition metal dichalcogenides can be intercalated with molecular cations to generate organic/inorganic superlattices which are stable in air.[20,30–40] This approach is particularly appealing to fabricating tailored superconducting circuits, as the selection of organic cations with different sizes enables fine-tuning of the material properties.[20,32] However, the emergence of superconductivity in organic-ion intercalated MoS$_2$ has not yet been reported.

In this work, we report on the emergence of superconductivity and CDW order in bulk crystals and exfoliated flakes of air-stable organic-ion intercalated MoS$_2$. In particular, MoS$_2$ intercalated with tetraethylammonium (TEA$^+$) features a CDW state and a superconducting transition with an onset around 3.9 K, leading to a fully developed zero-resistance state. Notably, the superconductive transition is recorded even after storing a crystal in the air for over six months. Conversely, in MoS$_2$ intercalated with cetyltrimethylammonium (CTA$^+$), a superconducting transition appears below 2.8 K, but the sample resistance remains finite even at the lowest temperature tested (1.9 K). Similarly, few-nm-thick flakes exfoliated from the TEA-intercalated MoS$_2$ crystals present a drop in the resistivity at 3.9 K but do not reach a zero-resistance state. These results indicate that the superconductivity in the organic-ion intercalated compounds emerges in spatially confined regions located in different atomic layers, which form a percolating superconducting path only in the case of bulk TEA-intercalated MoS$_2$ crystals. We ascribe this behavior to a spatially inhomogeneous distribution of the induced charge doping, which we probe at the nanoscale using scanning near-field optical microscopy.

## 2. Results and Discussion
### 2.1. Intercalation of MoS2 bulk crystals

The organic cations of the selected salts, (cetyltrimethylammonium (CTA$^+$) bromide and tetramethylammonium (TEA$^+$) bromide) are intercalated into MoS$_2$ bulk crystals with an electrochemical approach. We use a two-electrode setup in a galvanostatic configuration, i.e., driving a constant current between the cathode and the anode while monitoring the generated voltage (see **Figure 1**a, Methods section and Figure S1, Supporting Information). The electrolyte is composed of either CTAB or TEAB dissolved in acetonitrile (ACN). A MoS$_2$ crystal is anchored with indium to a platinum plate serving as the cathode and immersed in CTA$^+$ bromide or a TEA$^+$ bromide solution in ACN. At this electrode, the reduction half-reaction causes the intercalation of guest cations (G$^+$) species, driven by the electrostatic interaction between the negatively charged crystal and the positively charged molecules:

$$xe^- + xG^+(ACN) + MoS_2(s) \rightarrow G_X(MoS_2)(s)$$

Different materials are used as anode for the intercalation of the two organic cations, namely Ag for CTA$^+$ and Pt for TEA$^+$. The concurrent oxidation half-reactions are as follows: For CTA$^+$, the silver oxidizes and reacts with the bromide in solution precipitating in the form of AgBr on the surface of the anode, thus balancing the total electric charge of the system:

$$x\,Ag(s) + x\,Br^-(ACN) \rightarrow x\,AgBr\,(ACN) + x\,e^-$$

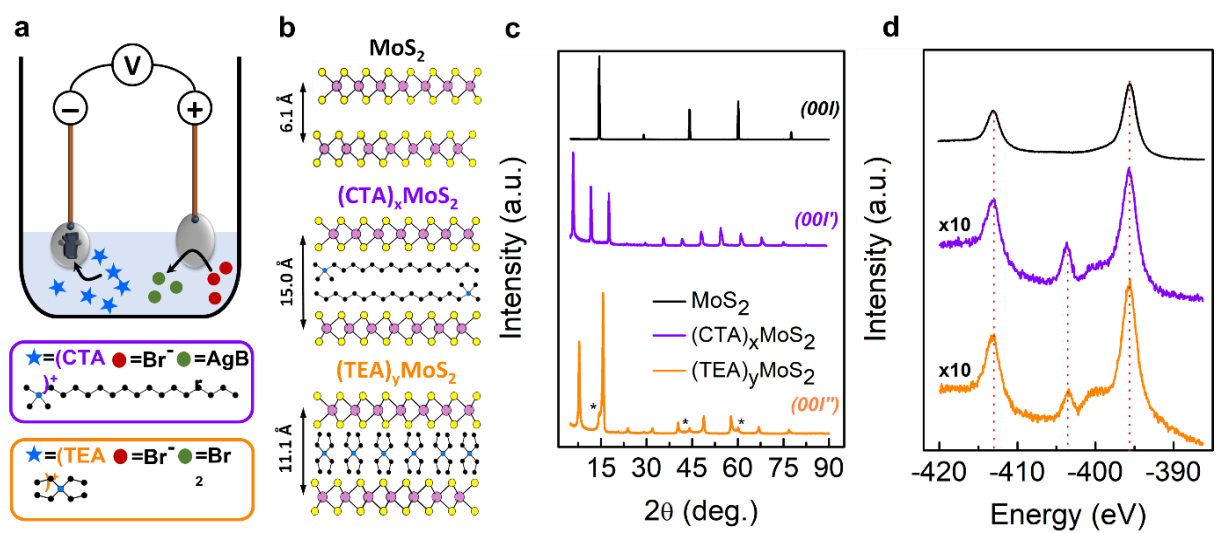

*Figure 1. Electrochemical intercalation of MoS$_2$. a) Schematic of the electrochemical cell used for the intercalation of cetyltrimethylammonium (CTA$^+$) and tetraethylammonium (TEA$^+$) ions in MoS$_2$. b) Crystal structure of MoS$_2$, (CTA)$_x$MoS$_2$ and (TEA)$_x$MoS$_2$ and corresponding interlayer separation, as extracted from the XRD diffraction patterns. c) XRD diffractograms of the pristine MoS$_2$ (black line), (CTA)$_x$MoS$_2$ (purple line) and (TEA)$_y$MoS$_2$ (orange line). Black stars denote the position of peaks corresponding to pristine MoS$_2$. d) Photoemission spectra of the Mo 3p and N 1s core levels for the pristine and intercalated compounds, MoS$_2$, (CTA)$_x$MoS$_2$, and (TEA)$_{yMoS2,}$ respectively.*

For the intercalation of TEA$^+$ ions, the oxidation of bromide at the platinum anode provides the electrons, generating bromine:

$$x\,Br^-(ACN) \rightarrow \frac{x}{2} Br_2(g) + x\,e^-$$

We cannot rule out secondary reactions involving other components of the cell such as the solvent taking place simultaneously to the main process. However, given the low current setting and the electrochemical stability of ACN we assume their impact is negligible.[41]

As a consequence of the intercalating process, the interlayer gap of the pristine host matrix expands (Figure 1b). This phenomenon allows us to assess the quality of the intercalation through X-ray diffraction (XRD). The diffractogram of pristine $MoS_2$ shows a collection of diffraction peaks arising from the reflections along the *(0 0 l)* direction (Figure 1c). After intercalation with both $CTA^+$ and $TEA^+$, the XRD pattern is profoundly modified, displaying different sets of peaks, *(0 0 l')* and *(0 0 l'')*. These signals are shifted to lower 2θ values compared to the pristine diffractogram, which according to Bragg's law indicates a larger interlayer distance (Figure S2, Supporting Information). In particular, the original vdW gap in pristine $MoS_2$ is 6.1 Å. Upon intercalation, this value extends to 15.0 Å for $(CTA)_xMoS_2$, and 11.1 Å for $(TEA)_yMoS_2$ intercalates, as depicted in Figure 1b. The difference in volumes of the organic compounds hosted in the $MoS_2$ lattice causes the two different interlayer distances. Interlayer variation Δd is equal to 5 Å for $(TEA)_yMoS_2$ and 8.9 Å for $(CTA)_xMoS_2$. The calculated Δd for $(CTA)_xMoS_2$ intercalates is compatible with the presence of two molecular layers in the van der Waals gap, where in each layer $CTA^+$ molecules lie parallel to the $MoS_2$ basal plane (d ~ 4.8 Å × 2, Figure 1b). In the case of the Δd calculated for $(TEA)_yMoS_2$, we propose that a monolayer of $TEA^+$ occupies the interlayer gap resulting in an interlayer distance of ~ 5.0 Å. We note that in the XRD pattern of $(TEA)_yMoS_2$ the peaks indicated by a black star correspond to the pristine $MoS_2$ phase, pointing out that the full intercalation of the crystal cannot be achieved. In fact, a more accurate XRD characterization indicates that the $TEA^+$ intercalation occurs in a region close to the external surfaces of the crystal, whereas the inner part of crystal remains pristine (Figure S3, Supporting Information).

We performed X-ray photoemission spectroscopy to understand the phenomena accompanying the intercalation at an elemental level. Figure 1d shows the photoemission spectra of the Mo 3p and N 1s core levels for the pristine and intercalated compounds. The intensities in the spectra of the intercalated samples are significantly lower than in the pristine crystal. We attribute these changes to the presence of a molecular layer on the surface of the flakes exposed after the exfoliation of the crystals. The position of the Mo 3p peaks does not significantly change after the intercalation, indicating that $(CTA)_xMoS_2$ and $(TEA)_yMoS_2$ retain the pristine trigonal prismatic phase (2H).[42–44] This is further confirmed by a micro-Raman characterization of the crystals, which shows no traces of the 1T/1T' phase (Figure S4, Supporting Information).

Additionally, the width of the Mo peaks in the XPS spectrum of the intercalates is 1.7 times larger than that of the pristine crystal, suggesting that the physicochemical environment of the Mo atoms is not homogeneous across the surface, producing slightly shifted components in the Mo peak which overall result in a larger width. The evolution of the Mo 3d and S 2p peaks (Figure S5 and Table S1, Supporting Information) further supports this interpretation of the results.

The analysis of the N1s peak shows two contributions to the N1s level in both $(CTA)_xMoS_2$ and $(TEA)_yMoS_2$. The relatively sharp peak located at a binding energy $E_b = 403.6$ eV corresponds to the intercalated molecules, the broader contribution at $E_b = 400$ eV most probably appears due to the presence of traces of acetonitrile remaining from the intercalating process.[45] The similar binding energy of the N1s peak in the two intercalated compounds can be explained based on that the very similar physico-chemical environment of the N atom the two molecules. From the integral of the N 1s component corresponding to the molecules, we can find that the number of intercalated molecules per unit area is approximately 40 % more in the case of $CTA^+$ than $TEA^+$ (Figure S6 and Table S2, Supporting Information). This can be explained considering that the long alkyl chains in $CTA^+$ are prone to form densely packed self-assembled architectures which fill very efficiently the vdW gap[46,47], while the more three-dimensional structure of $TEA^+$ generates a less densely packed assembly leaving more space for the co-intercalation of acetonitrile.

For the CTA-intercalated $MoS_2$, we estimate a stoichiometric molar ratio of $(CTA)_{0.28}MoS_2$ based on the gravimetric analysis, in good agreement with previous reports[20] (see Figure S1, Supporting Information). In the case of $TEA^+$, we cannot use these methods since the electrochemical process leaves traces of pristine unintercalated $MoS_2$. However, since our XPS characterization indicates the density of $TEA^+$ is 40 % lower than $CTA^+$, we use $(TEA)_{0.17}MoS_2$ as a rough approximation for the stoichiometry of the intercalated phase. These values indicate the average density of molecules in the intercalated phases, but we found that they significantly vary locally, as we will show further ahead in the manuscript.

## 2.2. Superconductivity and charge density wave in bulk MoS₂ crystals

The electrical transport properties of the two intercalated hybrids were measured using a 4-probe configuration. **Figure 2**a shows that, unlike pristine semiconducting $MoS_2$, $(CTA)_xMoS_2$ presents a metallic behavior characterized by an almost linear drop of the resistance, which reaches a plateau below 25 K. This temperature dependence of the resistance indicates that $(CTA)_xMoS_2$, which retains the structural 2H phase typically associated to the semiconducting state, becomes degenerately doped upon intercalation, behaving as a metal. A similar trend was recently observed in 2H MoS$_2$ intercalated with another organic compound, 1-ethyl-3-methylimidazolium.[32] Interestingly, we observe a marked decrease of the resistance below 2.8 K (Figure 2b), which we ascribe to the emergence of superconductive transition. However, the transition is rather broad, and the resistance does not reach zero in the probed temperature range. This behavior indicates that superconductivity only appears in spatially confined non-interconnected regions.

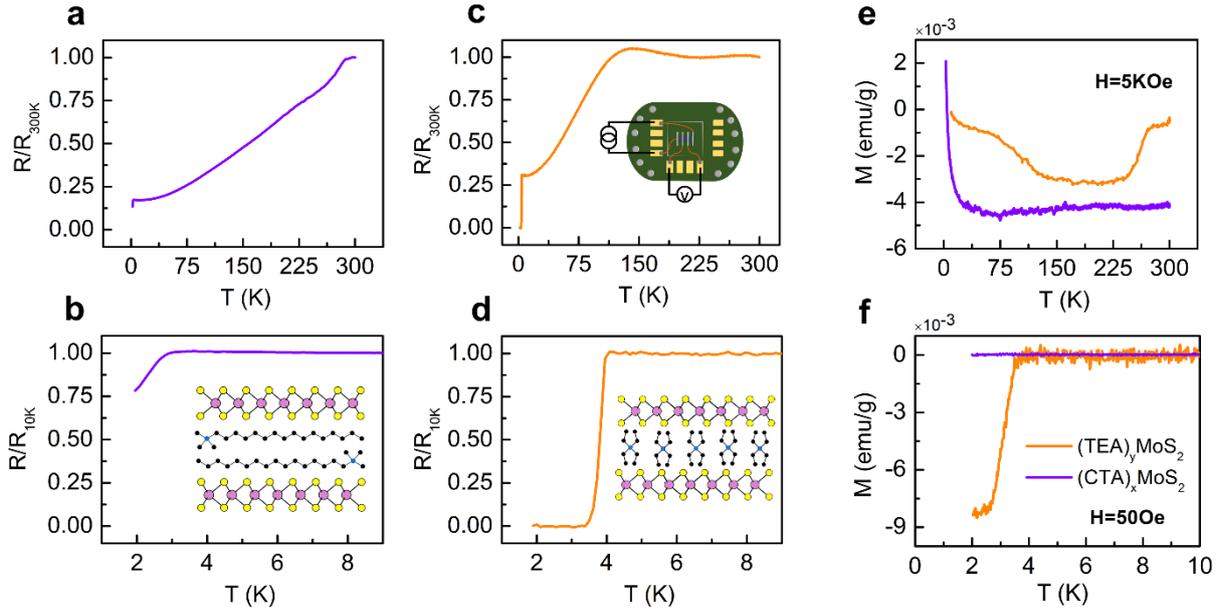

*Figure 2. Superconductivity in bulk crystals of organic-ion intercalated MoS$_2$. a) Four-probe resistance of bulk (CTA)$_x$MoS$_2$ as a function of the temperature. b) Resistance of (CTA)$_x$MoS$_2$ in the low-temperature range, showing an incomplete superconductive transition with an onset around 2.8 K. c) Four-probe resistance of bulk (TEA)$_y$MoS$_2$ as a function of the temperature. Inset shows a schematic of the sample holder and electrical contacts configuration used for all the bulk transport characterization. d) Resistance of (TEA)$_y$MoS$_2$ in the low-temperature range, showing a superconductive transition with an onset around 3.9 K. e) Magnetization M as a function of the temperature of both intercalates normalized to the moles of MoS$_2$. f) Magnetic response of both intercalates at low temperature (10 K-1.9 K). Meissner effect is only detectable in the case of (TEA)$_y$MoS$_2$.*

(TEA)$_y$MoS$_2$ shows markedly different behavior, featuring an increase of the resistance when lowering the temperature from 250 K to 135 K (Figure 2c). This observation is compatible with the presence of a CDW lattice distortion[37,48–50]. Moreover, (TEA)$_y$MoS$_2$ displays a superconducting transition below 3.9 K, reaching a well-defined zero-resistance final state (Figure 2d). This superconducting transition spans 400 mK, a value greater than the one found for perfect, homogenous superconductors[51]. This feature is a first indication that the intercalates are not homogenously doped and therefore different areas of the crystal become superconductive at different critical temperatures, thus broadening the transition of the bulk crystal.

Magnetization measurements above and below the critical temperature (Figure 2e and f) corroborate that $(TEA)_yMoS_2$ is characterized by a CDW ordering and undergoes a superconductive transition. While a featureless diamagnetic response is recorded for $(CTA)_xMoS_2$ in the temperature range between 300 K and 10 K under an applied field H = 5 KOe, an enhancement in the diamagnetic response is observed for $(TEA)_yMoS_2$ (Figure 2e) in a temperature range corresponding to the increase in the resistance in the transport measurements (Figure 2c). Similar behavior is observed in compounds displaying CDW order[49,50,52,53], reinforcing our interpretation of the emergence of CDW in $(TEA)_yMoS_2$.

In the 2 K – 10 K temperature range, we have measured the magnetization by applying a lower external field (H = 50 Oe), to minimally affect the superconductive transition. In this temperature range, we observe a sudden diamagnetic behavior for $(TEA)_yMoS_2$ below 3.9 K, which matches the superconducting transition observed in the resistance (Figure 2e). This feature is associated with the Meissner effect, which is a hallmark of superconductivity. The lack of this feature in $(CTA)_xMoS_2$ indicates that the volume of this crystal which turns superconductive is negligible, in agreement with the lack of a zero-resistance state at low temperatures.

We highlight that, whereas superconductivity has been previously reported in alkali metal intercalated $MoS_2$, to the best of our knowledge, it has not been found in $MoS_2$ intercalated with organic compounds.[32] Our data also shows that the interplay between molecular size, amount of charge, and other characteristics of organic ions such as flexibility of the carbon chains plays a fundamental role in determining the properties of the intercalated compounds. In particular, the different behavior of the two intercalated crystals correlates to the different doping and interlayer distance caused by the molecular guests and how they are arranged inside the van der Waals gap. In the present case, the charge carrier concentration in $(CTA)_xMoS_2$ and $(TEA)_yMoS_2$ can be calculated at respectively $3.3 \times 10^{14}$ and $2 \times 10^{14}$ electrons/cm$^{-2}$ per layer based on the estimated stoichiometry x = 0.28 and y = 0.17 (see also Figure S1, Supporting Information). Superconductivity in $MoS_2$ only emerges when the charge carrier density ($n_{2D}$) is between approximately $6 \times 10^{13}$ and $1 \times 10^{15}$ electrons/cm$^{-2}$ per layer, with the highest transition temperature measured at approximately $1.1 \times 10^{14}$ electrons/cm$^{-2}$ per layer.[14] Consequently, the estimated charge carrier concentration in (TEA)$_y$MoS$_2$ is closer to the optimal value for maximizing the critical temperature.

We highlight that the critical temperature in (TEA)$_y$MoS$_2$ is lower than in ionic-liquid gated compounds. This can be due to the superlattice crystal structure of the intercalates. In particular, the electrostatic screening in (TEA)$_y$MoS$_2$ is weaker than in ionic-gated crystals, since each layer is in contact with the insulating TEA$^+$ cations, and further apart from the conductive neighboring layers. In turn, the weaker electrostatic screening causes a larger Coulomb repulsion between the electrons in each layer, and, consequently, a weaker attraction in Cooper pairs. The same argument was proposed to account for the drop in the critical temperature of ionic-gated MoS$_2$ monolayers,[16] and it can also explain the vanishing superconducting state in (CTA)$_x$MoS$_2$. Due to its larger interlayer separation, the Coulomb repulsion in (CTA)$_x$MoS$_2$ is more pronounced than in (TEA)$_y$MoS$_2$, leading to an even weaker attraction of the Cooper pairs, and consequently a reduced critical temperature.

We emphasize that superconductivity in (TEA)$_y$MoS$_2$ is robust over time even for bulk crystals stored in air at room temperature with a relative humidity between 40% and 70%. A (TEA)$_y$MoS$_2$ crystal tested over a month shows that the superconducting phenomenon is not modified, with a critical temperature onset invariably pinned at 3.9 K (Figure S7, Supporting Information). Even crystals measured six months after intercalation displayed the superconductive transition to the fully developed zero resistance state. Further supporting the stability of the intercalates, the XRD features of (TEA)$_y$MoS$_2$ samples remain intact after six months in air (Figure S7, Supporting Information).

**2.3. Anisotropic Superconductivity in bulk (TEA)$_y$MoS$_2$**

Following the transport experiments, we performed an angular-dependent magnetoelectrical transport analysis to assess the anisotropy of the intercalate and to understand the dimensionality of the superconductivity. For this study, we focused on (TEA)$_y$MoS$_2$, which shows the well-defined zero-resistance state. **Figure 3**a and Figure 3b correspond to the evolution of the superconductive transition affected by increasing magnetic fields applied perpendicular or parallel to the basal plane of the crystal, respectively. As expected, due to the layered nature of the intercalates, the material displays a highly anisotropic behavior[54]. With the increased intensity of the applied magnetic field, the critical temperature lowers until the destruction of the Cooper-condensate phenomenon. When the magnetic field is applied perpendicular to the basal plane of (TEA)$_y$MoS$_2$, in the out-of-plane (OoP) configuration, superconductivity disappears at 0.5 T. The same effect is observed when the magnetic field is applied parallel to the layers of the intercalate, in the in-plane (IP) configuration. However, the required field to break the Cooper pairs for the IP configuration reaches 7 T. The disparity of the upper critical field values for IP and OoP configurations evidences the highly anisotropic nature of the intercalate, which is even higher than other layered superconductive materials such as $NbSe_2$[3]. Additionally, the IP critical field agrees with the Pauli paramagnetic limit $(1.86 \cdot T_c)$, indicating that the material behaves like a weakly-coupled superconductor following the Bardeen-Cooper-Schrieffer (BCS) theory.[55]

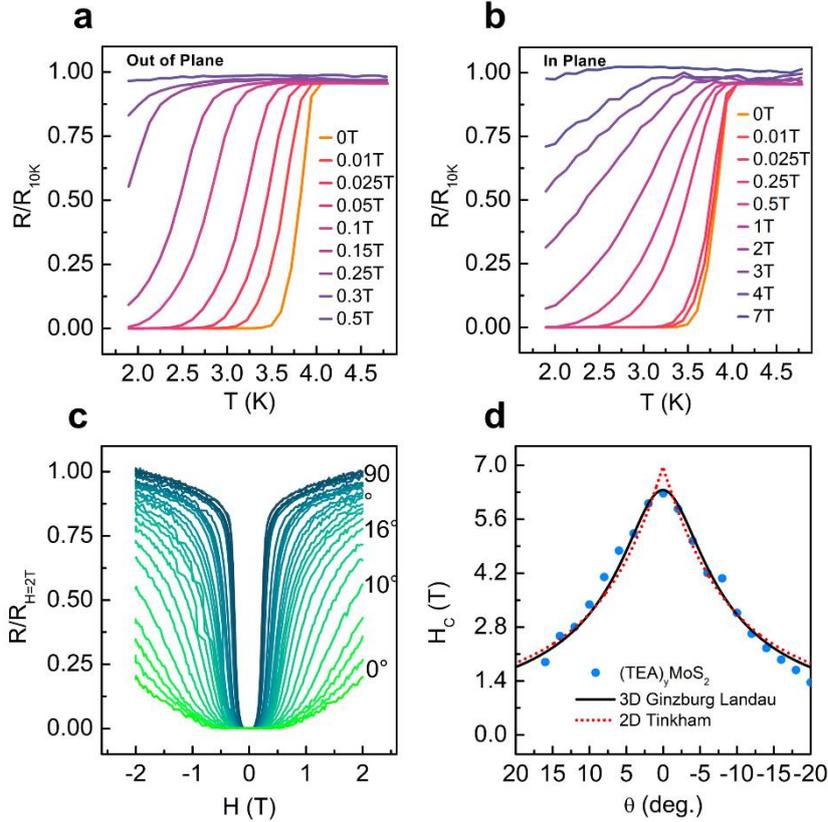

*Figure 3. Anisotropic Superconductivity in bulk $(TEA)_yMoS_2$. a) Effect of the out-of-plane magnetic field on the transition temperature of the $(TEA)_yMoS_2$ intercalate until total extinction of the superconductive behavior. b) Effect of the in-plane magnetic field on the transition temperature of the $(TEA)_yMoS_2$ intercalate until total extinction of the superconductive behavior. The curves are normalized to the 4-probe resistance value at 9K without a magnetic field. c) Magnetic field dependence of the four-probe resistance measured at 1.9 K at different angles between the field and the basal plane of the crystal. d) 3D Ginzburg-Landau and 2D Tinkham fitting of the upper critical field as a function of the magnetic field angle.*

Additionally, we recorded the crystal resistance as a function of a magnetic field applied at different angles (Figure 3c), and extracted the critical field $H_c$ for each configuration (Figure

S8, Supporting Information). In agreement with Figure 3 a and b, we observe a higher $H_c$ for low angles with a maximum at 0 degrees, when the magnetic field is parallel to the basal plane of the sample (Figure 3d). To assess the dimensionality of the superconductivity, we fitted the angular dependence of the $H_c$ to the Tinkham and to the Ginzburg-Landau (3D GL) formulas, which apply to a 2D and an anisotropic 3D superconductor, respectively (see Methods section). Our system is better described by the 3D GL model (black line in Figure 3d) indicating that the superconductive paths in $(TEA)_yMoS_2$ extend in the third dimension across different layers.

### 2.4. Superconductivity in exfoliated flakes of $(TEA)_yMoS_2$

After probing the superconductivity in bulk intercalated $MoS_2$, we addressed these properties in few-nm-thick flakes. This experiment is crucial to verify the potential impact of organic-ion intercalated $MoS_2$ for all-2D superconducting circuits for quantum computing.

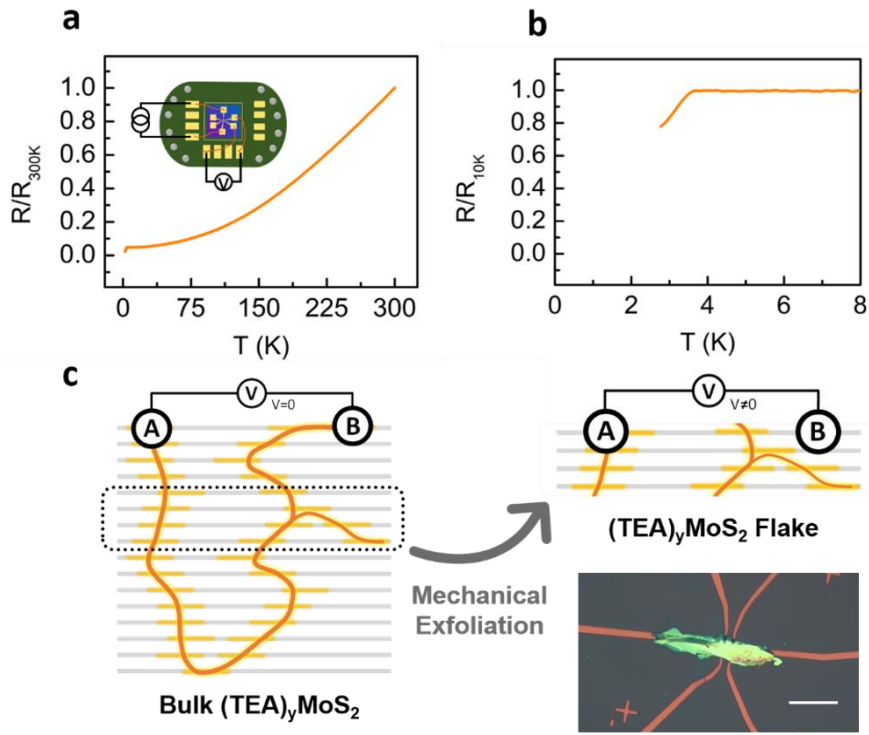

*Figure 4. Suppression of superconductivity in exfoliated flakes of $(TEA)_yMoS_2$. a) Evolution of the four-probe resistance as a function of the temperature of a 45-nm-thick $MoS_2$ flake stamped onto prepatterned contacts. Inset shows a schematic of the sample holder and electrical contact configuration used for the contacted flake. b) Incomplete superconductive transition of $(TEA)_yMoS_2$ c) schematic of the proposed percolating path model before and after exfoliation. A percolative superconductive path (shown in orange) is severed by the mechanical exfoliation, leading to a finite resistance state. An optical microscope image shows the mechanically cleaved $(TEA)_yMoS_2$ flake on the prepatterned contacts. Scale bar: 20 μm.*

To this end, we exfoliated a (TEA)$_y$MoS$_2$ crystal and dry-transferred it to prepatterned gold contacts with a Hall-bar-like configuration (inset **Figure 4**). We highlight that the occupation of the vdW gap in layered materials often results in ionic crystals that cannot be mechanically exfoliated, hindering the use of intercalated compounds in any micro-device application. We found that (TEA)$_y$MoS$_2$ remains exfoliable after intercalation, and it can be cleaved to obtain few-nm-thick flakes that can be stamped onto prepatterned contacts and measured electrically. The exfoliated flakes of (TEA)$_y$MoS$_2$ show a metallic behavior as temperature drops (Figure 4a and Figure 4b), following a similar trend as the bulk intercalate. However, the temperature dependence of the resistance did not show the CDW feature, which is present in all bulk intercalates. At low temperatures, the micro-flakes display an onset of the superconductive transition at a temperature similar to bulk intercalates (3.9 K) that does not lead to a zero-resistance superconductive state (Figure 4b), as the superconducting transition is less steep than the one observed in the bulk samples. We highlight that the temperature dependence of the resistance measured in the different flakes present the same qualitative behavior, with the onset of the resistance drop reproducibly at 3.9 K (Figure S9, Supporting Information).

To account for the different behavior of bulk crystals and exfoliated flakes, we propose that the superconductivity in intercalated MoS$_2$ does not extend over the whole crystal but rather to nanoscale regions. Therefore, a zero-resistance state is only possible if several superconducting areas in adjacent layers overlap across the crystal, defining a percolating path that extends in the direction perpendicular to the MoS$_2$ basal plane (Figure 4c). We note that this agrees with the magnetoelectrical transport measurements, which indicate a 3D superconductivity. As flakes are mechanically cleaved from the intercalated crystal, the probability for the electrons to find a complete superconductive path decreases. Consequently, the incomplete resistance drop observed in flakes corresponds to the emergence of superconductivity in spatially confined regions connected through resistive paths. We also highlight that a slightly different temperature dependence of the resistance is recorded when using different contact configurations in flakes contacted with multiple electrodes (Figure S9, Supporting Information), further indicating that the superconductive regions are not homogeneously distributed in each flake.

**2.5 Nanoscale optical conductivity in (TEA)$_y$MoS$_2$**

Since superconductivity in MoS$_2$ is directly related to the charge carrier density $n$,[14] the emergence of superconductivity in spatially confined regions can be understood as the result of a non-homogeneously distributed doping in the intercalated crystals. To characterize the homogeneity of the doping level, we probe the local conductivity σ (ω) of intercalated and pristine MoS$_2$ flakes using scattering-type scanning near-field optical microscopy (s-SNOM) and nanoscale Fourier-transform infrared nanospectroscopy (nano-FTIR).[56,57] These two methods are based on elastic light scattering at an atomic force microscope (AFM) tip, illuminated with monochromatic (s-SNOM) or broadband (nano-FTIR) laser illumination. The AFM tip acts as an antenna transforming the illuminating field into a strongly concentrated near field at the apex of the tip (nanofocus), creating a local excitation source for molecular vibrations, plasmons, or phonons at the surface of the sample[56,57] (**Figure 5**a and Methods section). In s-SNOM, interferometric recording of the tip-scattered field as a function of sample position yields images with a nanoscale spatial resolution of the amplitude *s* and phase *φ* of the complex near-field σ at the frequency of the illuminating light, simultaneously to the topography image. In nano-FTIR, asymmetric Fourier-transform spectroscopy of the tip scattered IR field provides nanoscale-resolved near-field amplitude s (ω) and phase φ (ω) in the spectral range of the broadband laser.

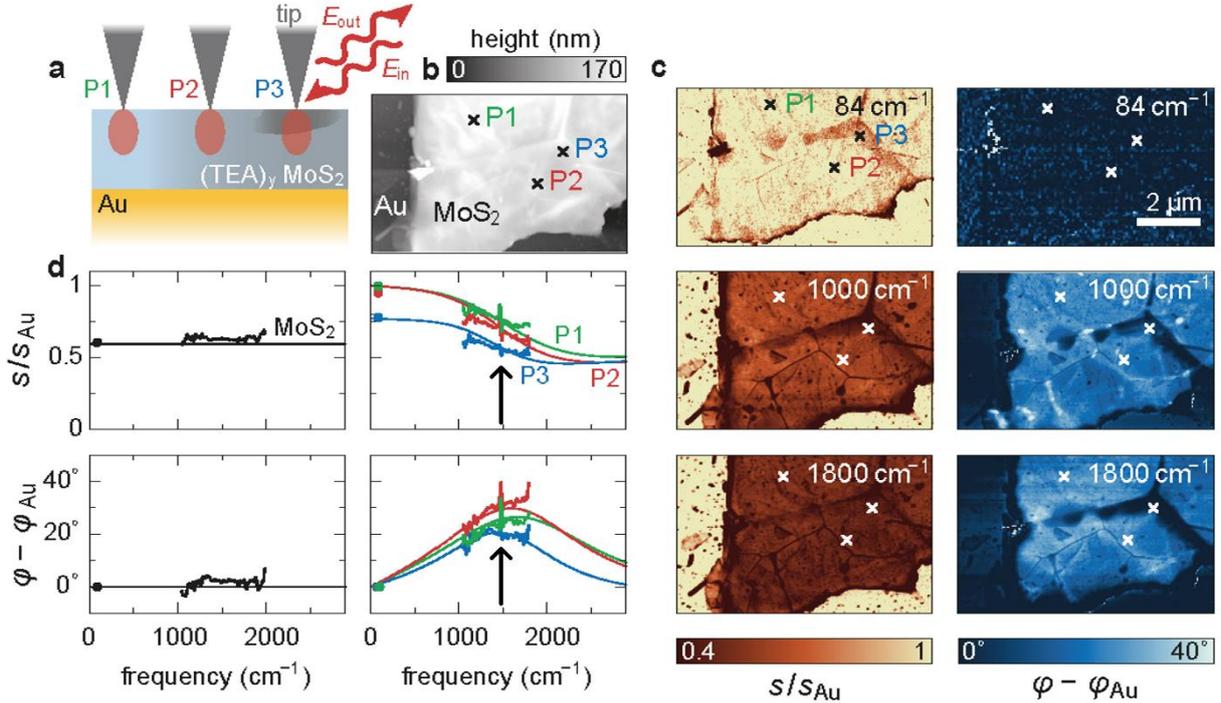

*Figure 5. Local conductivity mapping by THz and IR s-SNOM. (a) Illustration of local near-field optical probing by s-SNOM. A metal AFM tip is illuminated with IR or THz radiation, yielding a highly concentrated near-field spot (nanofocus) below the tip apex (red elliptical area). Near-field interaction between tip and sample modifies the tip-scattered field depending on the local sample properties. Detecting the tip-scattered field subsequently provides*

*nanoscale-resolved THz or IR images of the sample. Blue and grey regions within the intercalated (TEA)$_y$MoS$_2$ sample indicate highly and weakly doped (conductive) areas. (b) AFM topography image of a 150 nm thick (TEA)$_y$MoS$_2$ flake on a gold substrate exfoliated from an intercalated bulk crystal. (c) THz and IR amplitude (left) and phase (right) images were recorded at 84 cm$^{-1}$, 1000 cm$^{-1}$, and 1800 cm$^{-1}$ (top to bottom), normalized to the respective average values on the gold substrate. (d) Measured and calculated nano-FTIR spectra of pristine MoS$_2$ (left) and at positions P$_1$ to P$_3$ (indicated in panels b and c) on the intercalated (TEA)$_y$MoS$_2$ flake (right). The dots represent normalized amplitude and phase values obtained from the THz image (84 cm$^{-1}$) shown in panel c. The arrows highlight the peak corresponding to the TEA vibrational mode.*

Figure 5b shows the topographic image of a representative (TEA)$_y$MoS$_2$ flake exfoliated from a bulk sample and transferred onto a gold substrate, which is used as a metal reference within individual near-field images. s-SNOM imaging of the same flake shows no amplitude and phase difference in comparison to the gold substrate in the THz spectral range ($\nu$ = 84 cm$^{-1}$, Figure 5c), indicating a metallic behavior. In the IR region ($\nu$ = 1000 and 1800 cm$^{-1}$), the amplitude signals of (TEA)$_y$MoS$_2$ reduce significantly with increasing frequency, and the phase contrast increases (Figure 5c). Such near-field spectral behavior resembles a Drude response of highly doped semiconductors with plasma frequencies in the mid-IR spectral range.[58,59] However, the amplitude and phase images of the intercalated MoS$_2$ flakes are not homogeneous, indicating a spatial variation of the local conductivity, i.e., the carrier concentration. For comparison, flakes of pristine MoS$_2$ show a reduced near-field amplitude compared to the gold surface used as reference and the absence of phase contrast, indicating a significantly lower charge carrier concentration (Figure S10, Supporting Information).

To assess the degree of inhomogeneity in (TEA)$_y$MoS$_2$, we measure nano-FTIR spectra in exfoliated flakes of pristine and intercalated MoS$_2$. Pristine MoS$_2$ shows the flat nano-FTIR amplitude and phase spectra, characteristic of a semiconductor with low charge carrier doping (Figure 5d, left). For (TEA)$_y$MoS$_2$, we recorded the spectra at three representative positions marked as P$_1$, P$_2$, and P$_3$ (Figure 5d, right). In good agreement with the s-SNOM images, we observe decreasing amplitude signals with increasing wavenumber and large phase, indicating a much higher charge carrier concentration than in the pristine flake. Moreover, the spectra measured at the different points in (TEA)$_y$MoS$_2$ differ significantly from each other, supporting our conclusions regarding the probable inhomogeneous nature of the sample.

To verify that the nano-FTIR spectra indeed correspond to a Drude response and to determine the local conductivity of the $MoS_2$ flake, we theoretically reproduced the near-field spectra (see Methods Section). While at positions $P_1$ and $P_2$, we can accurately reproduce the experimental nano-FTIR spectra assuming a homogeneous carrier concentration in the flake (red and green curves in Figure 5d), at $P_3$ we must assume a 12-nm-thick undoped $MoS_2$ layer on top of doped $MoS_2$ to match the experimental data (blue curve in Figure 5d and schematics $P_3$ in Figure 5a. Therefore, our measurements indicate that the $(TEA)_yMoS_2$ comprises simultaneously undoped, weakly, and highly doped areas. From our model, we can extract the DC conductivity as the zero-frequency limit of $\sigma(\omega)$ (see Methods). We find $\sigma_{DC}$ = 499 Scm$^{-1}$, 497 Scm$^{-1}$, and 663 Scm$^{-1}$ for positions $P_1$, $P_2$, and $P_3$, respectively, which agree reasonably well with the typical value $\sigma_{DC}$ = 370 Scm$^{-1}$ obtained from the DC transport measurements of exfoliated flakes.

Finally, we highlight that the nano-FTIR spectra present a narrow spectral feature at 1479 cm$^{-1}$ in Figure 5d, which coincides with a molecular vibration of high oscillatory strength (H-C-H bending) of the intercalated TEA$^+$ molecule. Interestingly, the strength of the molecular spectral feature depends on the sample position, indicating an inhomogeneous distribution of the molecules. Importantly, the lowest intensity in the molecular vibrational features is recorded at P3, which corresponds to the position of the 12-nm-thick undoped region. Therefore, the existence of such undoped region can be ascribed to the local depletion of molecular dopants. In this way, our data allow us to relate the presence of the organic cation to the local doping level.

## 3. Conclusion

In conclusion, we have demonstrated the emergence of superconductivity and CDW ordering in $MoS_2$ intercalated with different organic cations. In particular, $(TEA)_yMoS_2$ bulk crystals display signatures of CDW-ordering in the temperature dependence of electrical resistance and magnetization. Moreover, they show a superconductive transition with an onset at 3.9 K, leading to a well-defined zero resistance state below 3.5 K. This behavior differs significantly from what we observe in bulk $(CTA)_xMoS_2$ crystals, which display a finite resistance even at 1.8 K. Surprisingly, few-nm-thick $(TEA)_yMoS_2$ flakes exfoliated from the superconductive bulk crystals do not reach a zero-resistance state. We ascribe this behavior to a non-homogeneous distribution of charge carriers, which confines the superconductivity in three-dimensional paths that can be severed by micromechanical exfoliation. We visualize the inhomogeneous charge carrier density at the nanoscale through SNOM and nano-FTIR. Our work shows that these techniques are ideal tools to simultaneously map local conductivity and molecular vibrations, thus allowing for correlating the presence of molecules with the electrical response of the system. Our discovery of molecule-dependent correlated electronic phases in organic-ion intercalated $MoS_2$ paves the way for the generation of novel hybrid material displaying tunable physical properties for cryo-electronics and quantum computing.

## 4. Experimental Section/Methods

*Materials:* $MoS_2$ crystals were purchased from HQ graphene. Acetonitrile was purchased from Sigma-Aldrich (anhydrous <0.001% $H_2O$). Cetyltrimethylammonium bromide (CTAB – purity >99%) was acquired from Acros Organics B.V.B.A. and tetramethylammonium bromide (TEAB - purity 98%) was purchased from Merck/Sigma-Aldrich. The platinum and silver electrodes were fabricated by pressing metallic pellets (Kurt Lesker, purity 99.99 %). Before usage, salts are dehydrated at 100° in vacuum (1 mbar) overnight. The electrodes need to be well polished and washed multiple times in acetone in an ultrasound bath followed by an isopropanol rinse.

*$MoS_2$ electrochemical intercalation:* The direct electrochemical intercalation of cetyltrimethylammonium ($CTA^+$) and tetramethylammonium ($TEA^+$) cations into $MoS_2$ bulk crystals takes place in a custom-made two-electrode cell, located in an $N_2$-filled glovebox ($H_2O$ < 0.1 ppm, $O_2$ < 0.1 ppm) at room temperature (23 °C). This cell consists of a glass vial that contains the electrolyte solution and two electrodes submerged in it. Two polymer-coated copper wires attached to a few-millimeter-thick metallic plate construct the electrodes. In the case of $CTA^+$ intercalation, a cetyltrimethylammonium bromide (CTAB) solution in acetonitrile (ACN) serves as electrolyte ([CTAB] = 80 mg $mL^{-1}$), and a silver plate serves as the anode, and a platinum plate as the cathode, where the $MoS_2$ is attached. Depending on the process, we can choose the most suitable metal to ensure that the production of damaging by-products that could interfere with the intercalating process does not occur. The typical mass of the $MoS_2$ crystals used for the electrochemical intercalation is in the range 2 mg – 5 mg.

We anchor the pristine crystal with indium and fully immerse it in the electrolyte. Then, to achieve the galvanostatic polarization of the electrodes a Keithley 2636 source meter drives a current of 30 μA between the two electrodes. In the case of $TEA^+$ intercalation, a tetraethylammonium bromide TEAB) solution in acetonitrile acts as electrolyte ([$TEA^+$] = 10 mg mL$^{-1}$). The cathode follows the same design used in the intercalation of CTAB. A platinum plate acts as the anode.

*MoS$_2$ X-Ray Diffractometry & Magnetometry:* X-ray diffraction is carried out with an XPERT-PRO diffractometer on bulk crystals in a ω/ω configuration using a copper cathode λ(Kα1) = 1.540598 as an X-ray source. Magnetization measurements as a function of the temperature were obtained using a physical properties measurement system (PPMS) in vibrating sample magnetometer mode. Additionally, for the magnetization measurements we normalized the raw data to the mass of the pristine MoS$_2$ crystals (MoS$_2$ mass before intercalation) for each sample: 3.17 mg and 2.55 mg for (CTA)$_x$MoS$_2$ and (TEA)$_y$MoS$_2$, respectively.

*X-ray photoemission spectroscopy:* X-ray photoelectron spectroscopy has been carried out holding the sample at room temperature and illuminating it with monochromatized Al K$_\alpha$ light (hυ = 1486.6eV) from a microfocus setup (SPECS Focus 600). The excited photoelectrons were collected by a SPECS 150 hemispherical analyzer at emission and incidence angles of 40° and 60°, respectively. The overall experimental resolution was extracted from Fermi edge analysis of a reference gold sample and resulted in 0.4 eV. All crystals were exfoliated inside the XPS chamber before the measurements. We performed the measurements in 4 different regions of the crystal that provided consistent results.

*Micro-Raman spectroscopy:* Room temperature Raman spectroscopy characterization was carried out in a Renishaw® inVia Qontor micro-Raman instrument equipped with a 100× objective using 633 nm laser as excitation source (diffraction gratings of 1800 l mm$^{-1}$ & 2400 l mm$^{-1}$) and an incident power <1mW to avoid damage of the sample during the measurements.

*Fabrication of devices based on exfoliated flakes:* Flakes were exfoliated from bulk crystals using the scotch tape (Nitto® SPV224P) and transferred to a Si/SiO$_2$ substrate prepatterned with Ti/Au contacts, using the dry polymer (Polydimethylsiloxane, PDMS) transfer technique. To ensure that most of the flakes on the tape were fully intercalated, we first isolated the top layer of the bulk intercalated crystal and then proceeded with the exfoliation of the same using the blue tape. The resulting flakes were transferred to the PDMS block and then aligned and stamped on the prepatterned electrodes a delamination-stamping system hosted in an Ar-filled glovebox.

*Magnetoelectrical transport measurements:* The electrical and magnetoelectrical transport data was obtained by contacting the bulk intercalated crystals manually to a chip carrier as shown in Figure 11, Supporting Information. The samples were then cooled down to 1.9 K using a PPMS (Quantum Design) while recording the evolution of the longitudinal resistance as a function of temperature and magnetic field. The applied current was provided by a Keithley 6221, and the voltage signal was detected by a nanovoltmeter Keithley 2182.

*2D Tinkham & 3D Ginzburg Landau Fits:* To fit our data to the 3D/2D models, we recorded the upper critical field ($H_c$) as a function of the angle of the magnetic field sweeping between 0 and 90 degrees with respect to the crystal sample. Then, we fitted the data to the 2D Tinkham (i) and the anisotropic 3D Ginzburg-Landau (3D GL) (ii) formulas using a least square method

$$(i) \quad \left(\frac{H_c \cdot \theta \sin(\theta)}{H_c^{ab}}\right)^2 + \left|\frac{H_c \cdot \theta \cdot \sin(\theta)}{H_c^c}\right| = 1$$

$$(ii) \quad \left(\frac{H_c \cdot \cos(\theta)}{H_c^c}\right)^2 + \left(\frac{H_c \cdot \sin(\theta)}{H_c^{ab}}\right)^2 = 1$$

*s-SNOM and nano-FTIR spectroscopy:* We used a commercial s-SNOM/nano-FTIR instrument (neaSNOM from neaspec GmbH, Munich) and standard PtIr-coated AFM tips (Arrow NCPt, Nanoworld, apex radius of about 30 nm) as near-field probes. The AFM was operated in tapping mode (tip oscillation amplitude was around 55 nm and oscillation frequency around ω = 260 kHz). Unwanted background signals were suppressed by demodulating the detector signal at the 3$^{rd}$ harmonic of the tip oscillating frequency, 3Ω, yielding background-free and amplitude and phase images/spectra, s and φ, respectively, with nanoscale spatial resolution. All images and spectra were normalized to the respective near-field signals obtained on the gold substrate. For recording the s-SNOM images, the tip was illuminated with infrared light from a wavelength-tunable quantum cascade laser (QCL) tuned to 1000 cm$^{-1}$ and 1800 cm$^{-1}$, respectively. The tip-scattered light was recorded with an integrated pseudo-heterodyne interferometer, simultaneously with the topography of the sample.[60–62]

For recording the nano-FTIR spectra, the integrated broadband laser light source (from Toptica, Munich) of the instrument covering the IR range between 1000-2000 cm$^{-1}$ was used.

The THz near-field image was recorded with a custom-made s-SNOM, which is based on the basic neaSNOM module (AFM and tip-illumination optics). The tip (Arrow NCPt, Nanoworld, apex radius of about 30 nm) was illuminated with the radiation from a wavelength-tunable THz gas laser (SIFIR-50, Coherent Inc., USA) tuned to 84 cm$^{-1}$ (2.52 THz). The tip-scattered light was recorded by synthetic optical holography (SOH) and a cryogen-free bolometer (QMC Instruments), yielding amplitude and phase images simultaneously to topography.[63]

*Modeling and fitting of the nano-FTIR spectra:* To simulate the spectra obtained by nano-FTIR spectroscopy, we calculated near-field spectra with the finite dipole model (FDM), [64,65] where the AFM tip is approximated by a prolate spheroid that is sinusoidally oscillating above the sample surface. The spheroid is characterized by its major half-axis length of $L = 300$ nm and an apex radius of $R = 30$ nm. For the oscillation amplitude $A$, we used the experimental tapping amplitude of 55 nm. The empirical model parameter $g$ was set to $g = 0.7e^{0.06i}$.

At the sample positions $P_1$ and $P_2$, the intercalated $MoS_2$ flake was modeled as a semi-infinite bulk sample. Accordingly, the near-field spectra were calculated with the FDM for semi-infinite samples as described in the literature.[64] For the intercalated (doped) $MoS_2$ we assumed a dielectric function $\varepsilon(\nu)$ that describes a Drude metal, $\varepsilon = \varepsilon_\infty[1-\nu_p^2/(\nu^2+i\nu\gamma_p)]$, where $\nu_p$ is the screened plasma frequency in wavenumber, $\gamma_p$ the electronic damping, $\nu$ the wavenumber, and $\varepsilon_\infty$ the high-frequency permittivity. We used $\varepsilon_\infty = 15$, according to values found in the literature[66]. $\nu_p$ and $\gamma_p$ were chosen such that the calculated near-field spectra match the experimental nano-FTIR spectra.

At sample position $P_3$ we modeled the sample as an undoped $MoS_2$ layer on top of a semi-infinite doped (intercalated) bulk $MoS_2$. Accordingly, the near-field spectra were calculated with the FDM for layered samples.[64] For the undoped $MoS_2$ layer we assumed a constant dielectric value of $\varepsilon(\nu) = 15$. The thickness was one of the parameters that were adjusted to match the experimental nano-FTIR spectra. The semi-infinite doped $MoS_2$ was treated as before at positions P1 and P2.

The values obtained for $\nu_p$ and $\gamma_p$ at positions P1 to P3 are listed in Tab. 1 together with the DC conductivity that can be obtained according to $\sigma_{DC} = 2\pi c\varepsilon_0\varepsilon_\infty\nu_p^2\gamma_p^{-1}$

| Measurement | $\nu_p$ [cm$^{-1}$] | $\gamma_p$ [cm$^{-1}$] | $\sigma$ [S cm$^{-1}$] |
| --- | --- | --- | --- |
| P1 | 2100 | 2200 | 499 |
| P2 | 2000 | 2000 | 497 |
| P3 | 2000 | 1500 | 663 |

*Statistical Analysis*: Resistance vs temperature measurements including those performed under the effect of magnetic fields were normalized using two different temperatures. Data between 300 K and 10 K has been normalized to the value of the corresponding sample at 300K, whereas data at low temperature between 10 K and 1.9 K were normalized using the value of the resistance of each sample shown at 10 K. Magnetization measurements were normalized to the mass of the intercalated compound. All calculations were performed using the software OriginPro. Fitting of the 2D Tinkham & 3D Ginzburg Landau models were obtained using Python. No further correction, smoothing or treatment was conducted. We measured the transport properties of a total of 12 bulk $(TEA)_y MoS_2$ crystals and 2 bulk $(CTA)_x MoS_2$ crystals, and 8 exfoliated $(TEA)_y MoS_2$ devices. All bulk $(TEA)_y MoS_2$ crystals and 7 out of 8 exfoliated $(TEA)_y MoS_2$ crystals displayed the superconductive transition. All these bulk crystals displayed an XRD pattern characterized by the same sets of peaks displayed in Figure 1c.

**Data availability**

The data that support the findings of this study are available from the corresponding author upon reasonable request.

**Supporting Information**

Supporting Information is available from the Wiley Online Library or from the author.

**Acknowledgments**

This work was supported by "la Caixa" Foundation (ID 100010434), under the agreement LCF/BQ/PI19/11690017, by the Spanish MICINN under Projects PID2019-108153GA-I00, PID2021-128004NB-C21, PID2021-123949OB-I00, PID2021-122511OB-I00. This work was also supported through the FLAG-ERA grant MULTISPIN, by the Spanish MCIN/AEI with grant number PCI2021-122038-2A. B.M.-G. thanks Gipuzkoa Council (Spain) in the frame of the Gipuzkoa Fellows Program. This work was supported by CEX2020-001038-M /AEI /10.13039/501100011033 under the Maria de Maeztu Units of Excellence Program. I.N. acknowledges financial support by the German Research Foundation (DFG) under project no. 467576442. The authors also thank SGIker Medidas Magneticas Gipuzkoa (UPV/EHU/ ERDF, EU) for the technical and human support.

The table of contents

**Percolating Superconductivity in Air-Stable Organic-Ion Intercalated MoS$_2$**
*Jose M. Pereira, Daniel Tezze, Iris Niehues, Yaiza Asensio, Haozhe Yang, Lars Mester, Shu Chen, Felix Casanova, Alexander M. Bittner, Maider Ormaza, Frederik Schiller, Beatriz Martín-García, Rainer Hillenbrand, Luis E. Hueso[*], Marco Gobbi[*]*

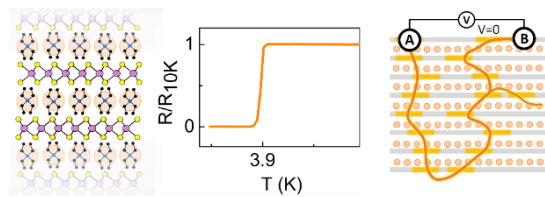

We report on the emergence of superconductivity in air-stable organic-ion intercalated MoS$_2$. Based on a multi-scale characterization, we conclude that superconductivity emerges in confined regions, leading to a zero-resistance state only when the intercalation creates a superconductive percolation path. The superconductive transition depends on the intercalated cation, evidencing the potential of organic-ion intercalation to tailor the properties of 2D materials.